\documentclass[aps,prl,twocolumn,floats]{revtex4}

\usepackage{amssymb,amsmath,graphicx}

\begin{document}

\title{Optimal cytoplasmatic density and flux balance model\\ under 
macromolecular crowding effects}

\author{Alexei Vazquez}

\affiliation{Department of Radiation Oncology, The Cancer Institute of New 
Jersey and UMDNJ-Robert Wood Johnson Medical School\\ 
195 Little Albany St, New Brunswick, NJ 08903, USA}

\date{\today}

\begin{abstract}

Macromolecules occupy between 34 and 44\% of the cell cytoplasm, about 
half the maximum packing density of spheres in three dimension. Yet, there 
is no clear understanding of what is special about this value. To address 
this fundamental question we investigate the effect of macromolecular 
crowding on cell metabolism. We develop a cell scale flux balance model 
capturing the main features of cell metabolism at different nutrient 
uptakes and macromolecular densities. Using this model we show there are 
two metabolic regimes at low and high nutrient uptakes. The latter regime 
is characterized by an optimal cytoplasmatic density where the increase of 
reaction rates by confinement and the decrease by diffusion slow-down 
balance.  More important, the predicted optimal density is in the range of 
the experimentally determined density of {\em E. coli}. We conclude that 
cells have evolved to a cytoplasmatic density resulting in the maximum 
metabolic rate given the nutrient availability and macromolecular crowding 
effects and report a flux balance model accounting for its effect.

\end{abstract}

\maketitle

\bibliographystyle{apsrev}

Macromolecular crowding affects the rate of biochemical reactions 
\cite{minton01}. It tends to increase reaction rates by increasing enzyme 
concentrations and to decrease reaction rates by reducing the diffusion 
coefficient of metabolites. The competition between these two factors 
results in a maximum reaction rate at intermediate crowding agent 
concentrations \cite{minton01}. Yet, it remains to be addressed whether 
this observation is true at the level of the whole cell metabolism. The 
macromolecular volume fraction of the {\it E. coli} cytoplasm is in the 
range 0.34-0.44 \cite{zimmerman91}, half the maximum packing density of 
spheres in three dimension. This observation indicates cell metabolism is 
operating under the crowding conditions created by its macromolecular 
components.

\begin{figure}

\centerline{\includegraphics[width=2.5in]{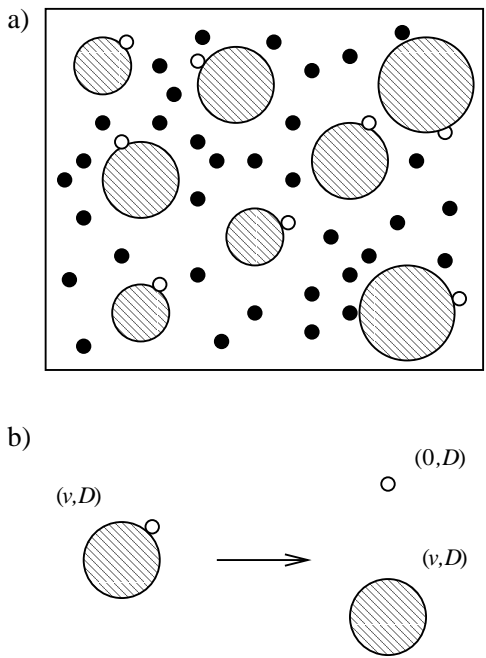}}

\caption{{\bf Metabolic model of the cytoplasm.} a) Schematic 
representation of the cytoplasm from the perspective of cell metabolism. 
The solid circles represent small metabolites. The dashed circles with 
attached small empty circles represent metabolic enzymes, the former 
representing their inert region and the latter the active site. b) We 
model an enzyme of molecular volume $v$ and diffusion coefficient $D$ as 
two different quasi-molecules. One representing the active site, with zero 
(or negligible) molecular volume and the same diffusion coefficient. The 
other representing the enzyme inert region, with same molecular volume and 
diffusion coefficient as the whole enzyme.}

\label{fig1}

\end{figure}

To investigate the effect of macromolecular crowding in the overall cell 
metabolism we focus on the model schematically represented in Fig. 
\ref{fig1}. (i) From the perspective of cell metabolism alone the 
cytoplasm is composed of metabolites and enzymes, the latter including 
metabolic enzymes, ribosomes and any other macromolecule catalyzing 
metabolic/biosynthetic processes. (ii) Metabolites are relatively small 
compared to enzymes and we neglect their contribution to crowding. (iii) 
The active site of most metabolic enzymes is relatively small compared to 
the whole enzyme. For all practical purposes the inert enzyme region is 
equivalent to an inert crowding agent. The inert region reduces the volume 
available to all solutes and, from a point of view of diffusion, 
collisions between metabolites and the inert region of enzymes are 
equivalent to collisions between the metabolites and inert crowding 
agents. Therefore, we model the contribution of active sites and the inert 
mass independently. We assume that each enzyme molecule contributes as two 
different fictitious quasi-molecules: one representing the enzyme active 
site (or union of active sites) and the other the inert enzyme region 
(Fig. \ref{fig1}b). The active site quasi-molecule is assumed to occupy a 
relatively small volume and its contribution to crowding is neglected. Its 
diffusion coefficient is assumed to be, however, the same as that for the 
original enzyme. The quasi-molecule representing the inert enzyme region 
is modeled as an inert crowding agent, with a size and diffusion 
coefficient equal to that of the corresponding enzyme. Because of their 
relatively larger sizes, enzymes have smaller diffusion coefficients than 
metabolites. We thus approximate the relative diffusion coefficient 
between enzymes and metabolites by the diffusion coefficient of the 
metabolites.

(iv) In a crowded media concentrations are effectively higher because part 
of the volume is occupied by the crowding agents. The ratio between the 
effective concentration $C$ and the concentration in an ideal solution 
$C_0$ is denoted by the activity coefficient $\gamma_C=C/C_0$. When the 
solute interacts with the crowding agents exclusively via steric repulsion 
$\gamma_C=V/V_a$ \cite{lebowitz65}, where $V$ is the total volume and 
$V_a$ is the volume available to the solute. The volume available to a 
solute can be approximated by the total volume minus the volume occupied 
by all crowding agents, resulting in

\begin{equation}\label{gamma}
\gamma_C = \frac{1}{1-v}\ ,
\end{equation}

\noindent where $v$ is the macromolecular volume fraction.

(v) The diffusion coefficient of a trace particle in a crowding media is 
also affected by the concentration of crowding agents. In general the 
diffusion coefficient is given by $D=\gamma_D D_0$, where $D_0$ is the 
diffusion coefficient in aqueous solution and $\gamma_D$ is a correction 
factor. To quantify the impact of crowding on metabolites diffusion we use 
the empirical exponential law

\begin{equation}\label{D}
\gamma_D = e^{ - \alpha v}\ ,
\end{equation}

\noindent where the exponent $\alpha$ is an empirical parameter 
\cite{furukawa91,kao93,dauty04,dix08}. The closest estimate for the 
cytoplasm comes from the experimental report $\alpha=5.8$ for fibroblast 
cells \cite{kao93}.

(vi) The kinetic models describing the rate of biochemical reactions as a 
function of the concentration of reacting metabolites can be quite complex 
and are unknown for most metabolic reactions. For example, consider the 
simple case where a substrate S is irreversibly transformed into the 
product P catalyzed by the enzyme E, with concentrations $S$, $P$, and 
$E$, respectively. The rate of this reaction is given by the 
Michaelis-Menten model $R=k_2 S E / (K_M + S)$, where $k_2$ is the rate of 
the conversion of the intermediate enzyme-substrate complex ES into the 
product ES$\rightarrow$P, $K_M=(k_{-1}+k_2)/k_1$ is the Michaelis-Menten 
or half-saturation constant, and $k_{-1}$ and $k_1$ are the rate of the 
intermediate steps ES$\rightarrow$E+S and E+S$\rightarrow$ES, 
respectively. The step E+S$\rightarrow$ES is diffusion limited and 
$k_1=4\pi D a$, where $D$ is the substrate diffusion coefficient and $a$ 
is the effective size of the enzyme active site. {\it Diffusion limited 
regime:} When $S\ll K_M$ the reaction rate can be approximated by 
$R\approx [k_2/(k_{-1}+k_2)] k_1 S E \propto D E$ and therefore the 
overall reaction is diffusion limited. In this limit most active sites are 
free and the reaction rate is limited by the rate of encounter of the 
enzyme active site and the substrate. {\it Saturation regime:} When $S\gg 
K_M$ the reaction rate is approximated by $R\approx k_2 E$. In this case 
most active sites are occupied and the reaction rate is limited by the 
chemical step ES$\rightarrow$P. The situation becomes more complicated for 
reactions involving more than one substrate, because of reversibility and 
other factors. Nevertheless, in general these two limiting scenarios, 
diffusion limited and saturation persist. As a first approximation we 
therefore assume that biochemical reactions are divided into two groups: a 
set ${\cal L}$ of diffusion limited reactions and a set ${\cal S}$ of 
reactions at saturation, with reaction rates given by

\begin{equation}\label{Rmodel}
R_i = \left\{
\begin{array}{ll}
g_i D_i C_i\ , & {\rm for}\ i\in {\cal L}\\
k_i C_i & {\rm for}\ i\in {\cal S}
\end{array}
\right.
\end{equation}

\noindent where $g_i$ is a model parameter containing all other 
contributions in the diffusion limited regime (e.g. metabolite 
concentrations) and $k_i$ is the rate of reaction $i$ in the saturation 
limit.

(vi) At the level of cell metabolism we use a steady state or flux balance 
model \cite{edwards02}. In the steady state the consumption and production 
of each metabolite balance

\begin{equation}\label{FB}
\sum_i S_{ji} R_i = 0\ ,
\end{equation}

\noindent where $i=1,\ldots,n$ as above is an index over reactions, 
$j=1,\ldots,m$ is an index over the metabolites, and $S_{ji}$ is the the 
stoichiometric coefficient of metabolite $j$ in reaction $i$ 
\cite{edwards02}. To account for the potential existence of a limiting 
nutrient we label by $i=1$ a reaction representing the nutrient uptake and 
assume

\begin{equation}\label{CC}
R_1\leq U\ , 
\end{equation}

\noindent where $U$ denotes the maximum uptake of the limiting nutrient. 
We also label by $i=n$ the metabolic objective or biomass vector, an 
effective reaction with a nonzero stoichiometric coefficient for each 
metabolite the cell produces and magnitude given by its relative ratio. 
Thus $R_n$ is our measure of metabolic rate.

(vii) Given a flux distribution $R_i$ we now calculate the volume fraction 
occupied by enzymes. The metabolic enzymes are characterized by their 
concentration $C_{0i}$, occupied volume fraction $v_i$, molar mass 
$\mu_i$, specific volume $\nu_i$ and kinetic model (\ref{Rmodel}). The 
occupied volume fraction is related to the enzyme concentration through 
the equation $v_i=\mu_i\nu_i C_{0i}$. In turn the enzyme concentration is 
related to the reaction rate through (\ref{Rmodel}). Putting these two 
relationships together, recalling that $C_i=\gamma_C C_{0i}$ and 
$D_i=\gamma_D D_{0i}$, and using (\ref{gamma}) and (\ref{D}), we obtain

\begin{equation}\label{vi}
v_i =(1-v) R_i\times\left\{
\begin{array}{ll}
\frac{\mu_i\nu_i}{g_iD_{0i}}e^{\alpha v}\ , & {\rm for}\ i\in{\cal L}\\
\frac{\mu_i\nu_i}{k_i}\ , & {\rm for}\ i\in{\cal S}\ .
\end{array}
\right.
\end{equation}

\begin{figure}

\centerline{\includegraphics[width=3in]{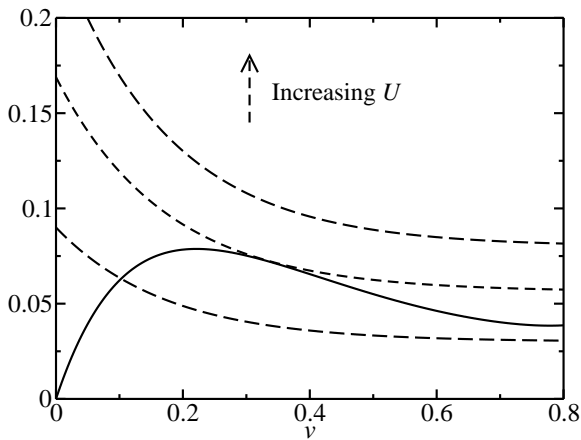}}

\caption{{\bf Calculation of the optimal volume fraction.} The left (solid 
line) and right (dashed lines) hand side of (\ref{VC}) for different 
values of $U$, assuming $S/L=2$ and $UL=0.03$, 0.057 and 0.08 from bottom 
to top. The intersection between these lines determines the optimal volume 
fraction $v^*$. For the upper dashed line, representing a larger $U$, 
there is no solution.}

\label{fig2}

\end{figure}

\noindent At this point we add up the volume fraction occupied by enzymes 
$v = \sum_i v_i$, which after some algebra results in

\begin{equation}\label{VC}
\frac{ve^{-\alpha v}}{1-v} = U \left[ L(r) + 
S(r) e^{-\alpha v} \right]\ ,
\end{equation}

\begin{equation}\label{LS}
L(r) = \sum_{i\in{\cal L}} \frac{\mu_i \nu_i}{g_i D_{0i}} r_i\ ,\ \ \ 
S(r) = { \sum_{i\in{\cal S}} \frac{\mu_i \nu_i}{k_i} r_i }\ .
\end{equation}

\noindent where $r_i=R_i/U$ are the reaction rates in units of $U$.

Under the model (i)-(vii), the optimal reaction rates $R_i$, 
$i=1,\ldots,n$, and volume fraction $v$ are obtained maximizing the 
metabolic objective $R_n$, subject to the flux balance (\ref{FB}), uptake 
capacity (\ref{CC}) and solvent capacity (\ref{VC})-(\ref{LS}) 
constraint. The solution of this optimization problem results in the 
following regimes.

\begin{figure}

\centerline{\includegraphics[width=3in]{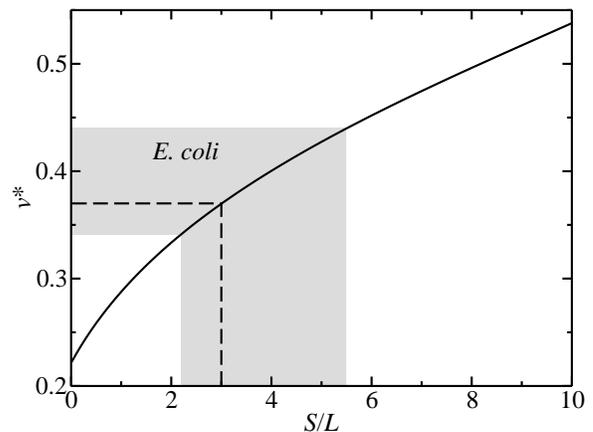}}

\caption{{\bf Optimal volume fraction.} The optimal volume fraction for 
$\alpha=5.8$ as a function of the ratio $S/L$ (solid line). The shadowed 
area was obtained using as input the volume fraction range 0.34-0.44 for 
{\it E. coli} \cite{zimmerman91} and computing the associated $S/L$ range 
from (\ref{VC}). The dashed line represents our prediction $v^*=0.37$ 
assuming $S/L=3$.}

\label{fig3}

\end{figure}

{\em Nutrient limited:} There is a threshold uptake capacity $U_c$ such 
that for $U<U_c$ the optimal reaction rates, denoted by $r^{(1)}_i$, are 
those maximizing the metabolic objective $R_n$ subject to the flux balance 
(\ref{FB}) and uptake capacity (\ref{CC}) constraint, $r^{(1)}_i$ are 
independent of $U$, and there is a volume fraction $v$ satisfying 
(\ref{VC}) with $L=L(r^{(1)})$ and $S=S(r^{(1)})$. The existence of this 
threshold uptake capacity is derived from the analysis of solutions to 
(\ref{VC}) with respect to $v$, given $L$ and $S$ (Fig. \ref{fig2}). The 
left hand side has a maximum at

\begin{equation}\label{v*}
v_0 = \left\{
\begin{array}{ll}
v_{m}\ , & 0<\alpha<4\\
\frac{1}{2} \left( 1 - \sqrt{ 1 - \frac{4}{\alpha} } \right)\ , & 
\alpha\geq 4\ .
\end{array}
\right.
\end{equation}

\noindent On the other hand the right hand side is a decreasing function 
of $v$, starting from $U(L+S)$ at $v=0$ and ending at $U(L+Se^{-v_m}$) at 
$v=v_m$, where $v_m$ denotes the maximum packing density of the 
cytoplasmatic components. For small values of $U$ the two curves intercept 
and (\ref{VC}) has a solution $v=v(UL,S/L)$, which is an increasing 
function of $U$. There is no solution, however, for large values of $U$. 
The maximum values of $U$ and $v$ where a solution exists, denoted by 
$U_c$ and $v^*$, are obtained when the tangents of the left and right hand 
side of (\ref{VC}) are equal as well. The simultaneous solution of these 
two equations results in an explicit relation between $U_c$ and $v^*$, and 
a transcendental equation for $v^*$ parametrized by $\alpha$ and 
$S/L$. For the intracellular value $\alpha=5.8$ we compute the 
maximum cell density as a function of $S/L$ (Fig. \ref{fig3}). 
It increases monotonically starting from $v_0\approx0.22$ (\ref{v*}) when 
all reactions are diffusion limited.

{\em Space limited:} For $U>U_c$ we cannot further increase the volume 
fraction beyond $v^*$ and (\ref{VC}) becomes a constraint on the reaction 
rates. In this regime the optimal reaction rates are obtained maximizing 
the metabolic objective $R_n$ subject to the flux balance (\ref{FB}), the 
uptake capacity (\ref{CC}), and the solvent capacity constraint

\begin{equation}\label{VC2}
\sum_i a_i R_i = 1\ ,
\end{equation}

\begin{equation}\label{a}
a_i = \left\{
\begin{array}{ll}
\frac{\mu_i\nu_i}{g_iD_{0i}} \frac{1-v^*}{v^*}e^{\alpha v^*}\ , & 
{\rm for}\ i\in L\\
\frac{\mu_i\nu_i}{k_i} \frac{1-v^*}{v^*}\ , &
{\rm for}\ i\in S\ .
\end{array}
\right.
\end{equation}

\noindent An equation similar to (\ref{VC2}) has been introduced before 
under the name of macromolecular crowding or solvent capacity constraint 
\cite{vazquez08a}. The coefficients $a_i$ have been named crowding 
coefficients as they quantify how much the reactions contribute to the 
crowding of the cytoplasm. Equation (\ref{a}) now introduces corrections 
to the previous calculations \cite{vazquez08a}, making a more precise 
accounting for the effect of macromolecular crowding on reactions rates.

To provide a quantitative prediction for the optimal volume fraction we 
need an estimate of $S/L$. Given that the kinetic parameters involved in 
these calculations are unknown for most biochemical reactions we cannot 
make a precise calculation at this point. Recent metabolite concentration 
measurements for {\it E. coli} \cite{bernett09} indicate that 83\% of the 
reactions have $S>K_M$, i.e. there are about three times more reactions in 
the saturation regime. Assuming that the values of $g_iD_{0i}$ and $k_i$ 
are not significantly different we would conclude that $S$ is about three 
times larger than $L$. In this case our calculations predict an optimal 
volume fraction of 0.37 (see Fig. \ref{fig3}). Experimental estimates of 
the macromolecular volume fraction of the {\it E. coli} cytoplasm indicate 
the lower and upper bounds 0.34 and 0.44 respectively \cite{zimmerman91}, 
containing the crude estimate 0.37 predicted above. A more conservative 
prediction is the interval between the lower bound 0.22 when all reactions 
are diffusion limited and the upper bound given by the maximum packing 
density $v_m$ when all reactions are at saturation. The latter can reach 
values as high as 0.8 when the particles have variable size \cite{sohn68}. 
This conservative range also contains the experimental report between 0.34 
and 0.44 for {\it E. coli} \cite{zimmerman91}.

In summary, there are two different metabolic regimes, nutrient limited 
and nutrient limited with a solvent capacity constraint. In the nutrient 
limited regime the metabolic rate is constrained by the stoichiometry of 
the set of biochemical reactions and the maximum uptake rate of the 
limited nutrient. In the nutrient limited with a solvent capacity 
constraint the metabolic rate is in addition constrained by macromolecular 
crowding effects. In particular there is an optimal macromolecular volume 
fraction resulting in the maximum metabolic objective. The optimal volume 
fraction is a function of the diffusion related exponent $\alpha$ and the 
ratio of reactions in a diffusion limited and saturation regimes. It 
interpolates between $v^*\approx0.22$ when all reactions are diffusion 
limited and the maximum packing density when all reactions are at 
saturation. Finally, we can write a general flux balance model that 
applies for both regimes: maximize the metabolic objective $R_n$ subject 
to the flux balance $\sum_iS_{ji}R_i=0$, uptake capacity $R_1\leq U$ and 
solvent capacity $\sum_i a_i R_i\leq1$ constraint. The latter inequality 
is satisfied with the less than sign for $U<U_c$ and with the equal sign 
for $U>U_c$.

We have made several assumptions that could affect the obtained results. 
The equations (\ref{gamma}) and (\ref{D}) characterizing the impact of 
macromolecular crowding on effective enzyme concentrations and metabolite 
diffusion coefficients may require further corrections when the occupied 
volume fraction gets closer to the maximum packing density. In the 
intracellular media the typical occupied volume fraction is about half the 
maximum packing density 0.74 of spheres in three dimensions. Therefore, we 
expect equations (\ref{gamma}) and (\ref{D}) to be sufficiently good 
approximations. The kinetic models cannot always be approximated by the 
extreme cases of diffusion limited and saturation regimes. As a 
consequence the plot in Fig. \ref{fig3} may be slightly different.

Taking together our results indicate that at high metabolic rates there is 
an optimal intracellular density where the increase of reaction rates by 
confinement and the decrease by diffusion slow-down balance. Since it is 
the optimal density resulting in the maximum metabolic rate, an increase 
of the density of enzymes beyond this optimal value will result in a 
decrease of the metabolic rate. Although this may sound counter intuitive, 
it follows from the fact that beyond the optimal density the slow-down 
of diffusion starts to dominate, diminishing the overall metabolic rate. 
More important, the experimentally determined density of {\it E. coli} is 
in the range predicted from our model. We thus claim that cells have 
evolved an intracellular density that results in maximum metabolic 
capabilities given the macromolecular crowding effects.

{\bf Acknowledgements:} Many thanks to The Simons Center for Systems 
Biology at the Institute for Advanced Study where part of this work was 
performed.


\end{document}